# Benefitting from the Grey Literature in Software Engineering Research

**Vahid Garousi**

Queen's University Belfast
Belfast, Northern Ireland, UK
v.garousi@qub.ac.uk

**Michael Felderer**

University of Innsbruck, Innsbruck, Austria & Blekinge Institute of Technology, Sweden
michael.felderer@uibk.ac.at

**Mika V. Mäntylä**

University of Oulu, Oulu, Finland
mika.mantyla@oulu.fi

**Austen Rainer**

Queen's University Belfast
Belfast, Northern Ireland, UK
a.rainer@qub.ac.uk

**Abstract:** Researchers generally place the most trust in peer-reviewed, published information, such as journals and conference papers. By contrast, software engineering (SE) practitioners typically do not have the time, access or expertise to review and benefit from such publications. As a result, practitioners are more likely to turn to other sources of information that they trust, e.g., trade magazines, online blog-posts, survey results or technical reports, collectively referred to as Grey Literature (GL). Furthermore, practitioners also share their ideas and experiences as GL, which can serve as a valuable data source for research. While GL itself is not a new topic in SE, using, benefitting and synthesizing knowledge from the GL in SE is a contemporary topic in empirical SE research and we are seeing that researchers are increasingly benefitting from the knowledge available within GL. The goal of this chapter is to provide an overview to GL in SE, together with insights on how SE researchers can effectively use and benefit from the knowledge and evidence available in the vast amount of GL.

# 1. Introduction

Scientists generally place the most trust in peer-reviewed, published information, such as journals and conference papers, according to Institute for Work & Health (2019). By contrast, software practitioners typically do not have access to peer-reviewed publications, or the time or expertise to read such publications. As a result, practitioners are more likely to turn to other sources of information that they trust, e.g., trade magazines, online blog-posts, question-answers sites, survey results or technical reports, collectively referred to as *Grey Literature (GL),* as mentioned in a technical report by the Institute for Work & Health (2019). Furthermore, practitioners also share their ideas and experiences as GL, which can serve as a valuable data source for research. Indeed, Devanbu et al. (2016) and Rainer et al. (2003) both found that practitioners most trust their peers, particularly local experts. This situation can lead to various negative outcomes for research, e.g., the limited quality and quantity of communication between researchers and practitioners as reported by Garousi et al. (2019), and the limited relevance and applicability of many research papers when applied into industrial settings.

In an online article by the University of New England (2019), it is mentioned that: "*Much grey literature is of high quality. Grey literature is often the best source of up-to-date research on certain topics, such as rural poverty*". We wonder about the comparable situation in software engineering (SE). As examples, there is a book by Brooks (1995), entitled *The Mythical Man-Month*, and also another book by DeMarco and Lister (2013), entitled *Peopleware*. While both of these books formally belong to the GL, since they are books, and, yet, they are highly-cited in SE research.

We and many researchers, e.g., Elliott (2019), share the opinion that: "*if used with care, grey literature can open up valuable additional sources of information for researchers*". Furthermore, according to a paper by Farace (1997), the growth rate of GL was 3-4 times that of conventional peer-reviewed literature. With the major advancement of the internet, we believe that the dissemination rate of GL would be much higher now.

Whilst GL can offer a wealth of additional information for researchers, some of this information being much more current than research, GL should also be treated with caution and cross-checked with other sources. For example, an assessment by the Intergovernmental Panel on Climate Change (IPCC) of climate science in 2007 was subsequently criticized by the Inter-Academy Council (IAC), an umbrella council for science academies. According to Rincon (2010), the IAC reported that part of the IPCC report contained statements based on little evidence, and the use of GL in that assessment "*sparked controversy*".

> There is a great potential for benefitting from grey literature in software-engineering research.

Other papers put forward bold ideas relating to GL, e.g., Banks (2006) suggested the notions of a "*continuum of scholarship*" and "*the eventual collapse of the* *distinction between grey and non-grey literature*", implying that different types of literature (peer-reviewed and grey) are, or could be, merging into each other.

Over many years, SE research has used a variety of practitioner-generated content in close collaboration with practitioners, e.g., the work by Molléri et al. (2016) on interviews, opinion surveys, project documents; and the work by Sharp et al. (2010) on ethnographies. Garousi et al. (2019) recently suggested the use of GL as a knowledge source in SE research. Papers such as the one by Garousi and Mäntylä (2016), and Williams and Rainer (2017). As reported by Garousi et al. (2019), a large number of SE practitioners write and share technical writings as GL, e.g., in the form of blog-like documents, videos, and white papers. As recent work in SE has shown – e.g., Garousi et al. (2016a), Rainer (2017), Williams and Rainer (2017), and Rainer and Williams (2019) – there is great potential for benefitting from GL in SE research.

Practitioners have shared their ideas and experiences online for many years and thus GL itself is not a recent topic in SE. However, using, benefitting and synthesizing knowledge from the GL in SE is a contemporary issue in SE research and in empirical SE. The goal of this chapter is to provide an overview to GL, together with insights into using and benefitting from the knowledge available in the vast amount of GL in SE research.

The remainder of this chapter is organized as follows. We first review the general concept of GL and provide further background information. We then review the state of GL in SE research, including context, types, diversity and scale of GL in SE research and practice. We then suggest and discuss a selection of approaches for using and analyzing GL in SE research.

## 2. The general concept of grey literature

In this section, we provide an overview to the general concept of GL, including different types of GL, and how GL has been conceived and used in other research disciplines.

### 2.1. What is GL and what are its types, in general?

Though there are many definitions of GL in the literature, they are quite similar. The Cochrane handbook for systematic reviews of interventions Lefebvre et al. (2008) defines GL as "*literature that is not formally published in sources such as books or journal articles*". According to Institute for Work & Health (2019) GL is essentially any document that has *not* gone through formal peer review for publication. There is an annual conference on GL[1] and an international journal on GL[2]. There is also a Grey Literature Network Service[3], which is "*dedicated to*

---

[1] www.textrelease.com
[2] http://www.greynet.org/thegreyjournal.html
[3] www.greynet.org

*research, publication, open access, education, and public awareness to grey literature*".

However, the which types of sources are considered as GL depends to some extent on the research discipline. Therefore, models to classify source types into categories of GL are very helpful. There are many models for classifying different categories of GL and GL sources, e.g., the model by Adams et al. (2016), shown in Figure 1, stems from the management sciences. This model has two dimensions: (1) expertise and (2) outlet control. Both dimensions run between the extremes "unknown" and "known". Expertise is the extent to which the authority and knowledge of the producer of the content can be determined. Outlet control is the extent to which content is produced, moderated or edited in conformance with explicit and transparent knowledge-creation criteria. Rather than having discrete bands, the gradation in both dimensions is on a continuous range between known and unknown, producing the shades of GL. To emphasize: the figure is not intended to suggest discrete boundaries between the tiers.

The model presented in Figure 1 is comparable with Table 1, developed by Giustini and Thompson (2010), that shows the spectrum of the 'white', 'grey' and 'black'. The 'white' literature is visible in both Fig. 1 and Table 1 and indicates that both expertise and outlet control are fully (or at least sufficiently) known. With the examples presented in Table 1, GL corresponds mainly to the $2^{nd}$ tier in Figure 1, with moderate outlet control and expertise.

'Black' literature are at the low-end of both the outlet control and credibility spectrums. Blogs, but also emails and tweets, mainly refer to ideas, concepts or thoughts that are not peer reviewed by any 'outlets'. They are typically in the $3^{rd}$ tier of the model presented in Figure 1.

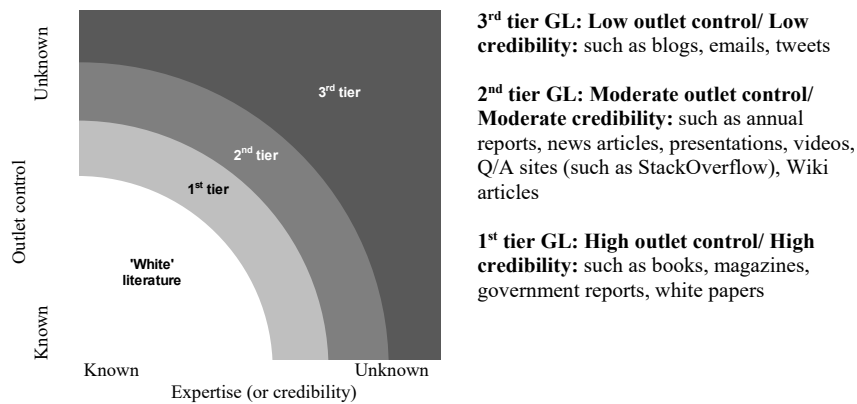

**Figure 1- "Shades" of grey literatures (based on Adams et al. (2016))**

We noted earlier that there are "shades" of grey in the classifications given in Fig. 1 and Table 1 and, depending on the degree of peer-review during the process of creating the item of GL, a specific item of GL can be in different tiers of Figure 1. For example, in a GL review study of micro-services, Soldani et al. (2018)

identified 20 blog posts for further analyses, using a quality checklist to identify the higher-quality GL. As a contrasting example that analyzed the types, frequencies, and findability of inter-disciplinary GL, Marsolek et al. (2018) treated conference papers as GL. This is because, in some disciplines, conferences accept all submitted papers with no peer review. However, in SE research at least, the highly ranked conferences have established peer review processes to select submitted papers for publication in a similar process to journals. Thus, the SE research community does not in general treat conference papers as GL.

> Grey literature sources can be classified according to the two dimensions: expertise and outlet control. Expertise is the extent to which the authority and knowledge of the producer of the content can be determined. Outlet control is the extent to which content is produced, moderated or edited in conformance with explicit and transparent knowledge-creation criteria.

As further contrasting examples, MSc and PhD theses are often reviewed by several examiners, and therefore are also often peer reviewed. Also, in most software companies who intend to share technical reports or white papers online, such documents are almost always reviewed to some degree by peers, and therefore such publications could be considered as peer-reviewed literature. The peer review process for technical documents differs to the peer review of academic publications, however. For example, academic peer-review is often done anonymously by reviewers who are independent, with less potential for conflicts of interest. The peer review process is also managed by an independent editor. By contrast, the peer review of technical documents, in practice, may often be undertaken by known colleagues. Thus, in summary, we conclude that what constitutes GL depends on the standards of the respective research discipline.

We also recognize that the rise of social media is increasing the extent of GL in SE. Storey et al. (2014) write of "The (R)Evolution of Social Media in Software Engineering". They illustrate that communication and social media produce data through different channels and this communication evolves over the years. For example, Usenet, Email List, and SourceForge used to be popular but currently tools such as Stack Overflow, Slack, and GitHub dominate. Social media is usually $3^{rd}$ level (tier) GL, in Fig. 1, but some of content sources such as Stack Overflow or Wikipedia can be considered $2^{nd}$ level GL as there are informal controls and other people can edit and improve the content. Williams and Rainer (2017) recommend that GL materials "*need to be rigorous, relevant, well written and experience based for them to be considered credible to [SE] researchers*". Another consideration is that as one lowers the quality threshold, i.e., move from tier 1 to 3 (in Figure 1), the amount of available literature grows to enable large-scale quantitative analysis. The scale of this GL is further addressed in Sect. 3.5.

**Table 1- Spectrum of the 'white', 'grey' and 'black' literature. From: Giustini and Thompson (2010)**

| 'White' literature | 'Grey' literature | 'Black' literature |
| --- | --- | --- |
| • Published journal papers<br>• Conference proceedings<br>• Books | • Preprints<br>• e-Prints<br>• Technical reports<br>• Lectures<br>• Data sets<br>• Audio-Video (AV) media<br>• Blogs | • Ideas<br>• Concepts<br>• Thoughts |

Due to the limited control of expertise and outlet in GL, it is important to also identify GL producers. Giustini and Thompson (2010) identified the following GL producers: (1) Government departments and agencies (i.e., in municipal, provincial, or national levels); (2) Non-profit economic and trade organizations; (3) Academic and research institutions; (4) Societies and political parties; (5) Libraries, museums, and archives; (6) Businesses and corporations; and (7) Freelance individuals, i.e., bloggers, consultants, and Web 2.0 enthusiasts. Marsolek et al. (2018) found that GL was present in the majority (68%) of the subject databases and almost all institutional repositories (95%).

## 2.2. GL in other research disciplines

GL is already established in a number of other research disciplines. Marsolek et al. (2018) examined 118 subject databases used by academic researchers, together with 115 repositories held by North American institutions, which included GL. The databases and repositories covered the arts, business, education, health sciences, humanities, multidisciplinary research, natural sciences, physical sciences and engineering, and social sciences.

> Grey literature is already established, as a source of knowledge / evidence, in many other research disciplines.

Luzi (2000) identified stages in the growth of GL, from its first appearance in the post-war period to its evolution into electronic GL, and analyses a selection of studies and conferences organized up to the 1990s. He also examines the first databases: these transformed the way in which GL was collected and distributed. Luzi's review is of course dated now by about 20 years. In contrast to Luzi's retrospective, Banks (2006) took a more prospective view and considers political and technological aspects for increasing access to valuable GL. For Banks, institutional repositories present an exciting opportunity for both the preservation and retrieval of GL.

Other relevant work includes discussions by Thompson (2001) of ways in which GL in engineering can be acquired and used, and arguments provided by McKimmie and Szurmak (2002) on how grey questions can drive research.

## 2.3. Usage and analysis of GL in the CS research

GL has also received attention in the computer science (CS) research community, e.g., computational linguistics, data and knowledge engineering, information retrieval, database, expert systems. Studies have analyzed the GL data to answer a variety of research questions. For example, Swanson et al. (2014) focused on identifying narrative clauses in personal stories. Their study used 50 personal stories drawn from 5,000 blog posts. Facca and Lanzi (2005) reported a survey on mining "interesting" knowledge from weblogs. Park et al. (2010) analyzed 588 sentences from 6,000 blog posts on WordPress. Kurashima et al. (2009) analyzed 29M blog posts collected using the BlogRanger 2.0 API. In another study, Kurashima et al. (2006) analyzed 62,396 articles from two Japanese blog hosting sites. Finally, Inui et al. (2008) analyzed 50M posts from 150M weblog posts (in Japanese). Bansal et al. (2007) developed *BlogScope*, a system for analyzing temporally ordered streaming text online. At the time, *BlogScope* "*... track[ed] more than 36 million blogs with more than 837 million posts in the blogosphere... [fetching on average] 14,000 new documents every hour.*"; a quote from Lakshmanan and Oberhofer (2010). The service was shut down in early 2012.

## 3. Grey literature in software engineering

In this section, we review the state of GL in SE, considering the context, types, diversity and scale of GL.

### 3.1. What is GL in SE?

Based on the two dimensions to classify GL defined in the previous section, i.e., expertise and outlet control, GL in SE can be defined is any material about SE that is not formally peer-reviewed nor formally published. We summarize the concept of GL in SE using a UML context diagram (see Figure 2).

> Grey literature in SE can be defined is any material about SE that is not formally peer-reviewed nor formally published.

At the center of this diagram lies the "*Technical information (writing)*" class which has two sub-classes: *Paper in academic literature*, and *Artefact in grey literature*. A *technical information* piece is written by one or more *Authors* and is read by one or more *Readers*, who are themselves *Software Engineers*. A *Software Engineer* has two sub-types: *Practitioner*, and *Researcher*. An *Artefact in grey literature* could be a *Blog-like document*, or *Video*, or *White-paper*, etc. According to Rainer (2017), an *Artefact in grey literature* could include *Conclusions* which in turn may be backed (supported) by *Reasoning*, *Experience*, and/or *Evidence*; and illustrated by one or more *Examples*. The *Practitioner* writes an *Artefact in grey literature* using her/his professional experience of software practice (empirical world).

### 3.2. Scale of the software engineering community: academia vs. industry

To further understand the role and position of GL in SE practice and SE research, we present a high-level view of the community of SE practice versus research. We look at the estimated population of the two communities.

According to a report by Evans Data Corporation (2018), there were about 23 million software developers worldwide in 2018, and that number is estimated to reach 27.7 million by 2023. According to an *IEEE Software* paper by Briand (2012), "*4,000 individuals" are "actively publishing in major [SE] journals*", which can be used as the estimated size (lower bound) of the SE research community. If we divide the two numbers, we can see that on average, there is one SE academic for every 5,750 practicing software engineers, indicating that the size of the SE research community is very small compared to the size of the SE practitioner community. We visualize the two communities and the current state of collaboration in Figure 3, which has been taken from the work of Garousi et al. (2018). Chapter 18 of this book also presents important concepts about industry-academic collaborations in software engineering.

As visualized in Figure 3, while there exist established ways to enable knowledge flow from software industry to academia, e.g., interviews, opinion surveys, ethnography, we believe that benefitting from GL materials is another prominent enabler for this.

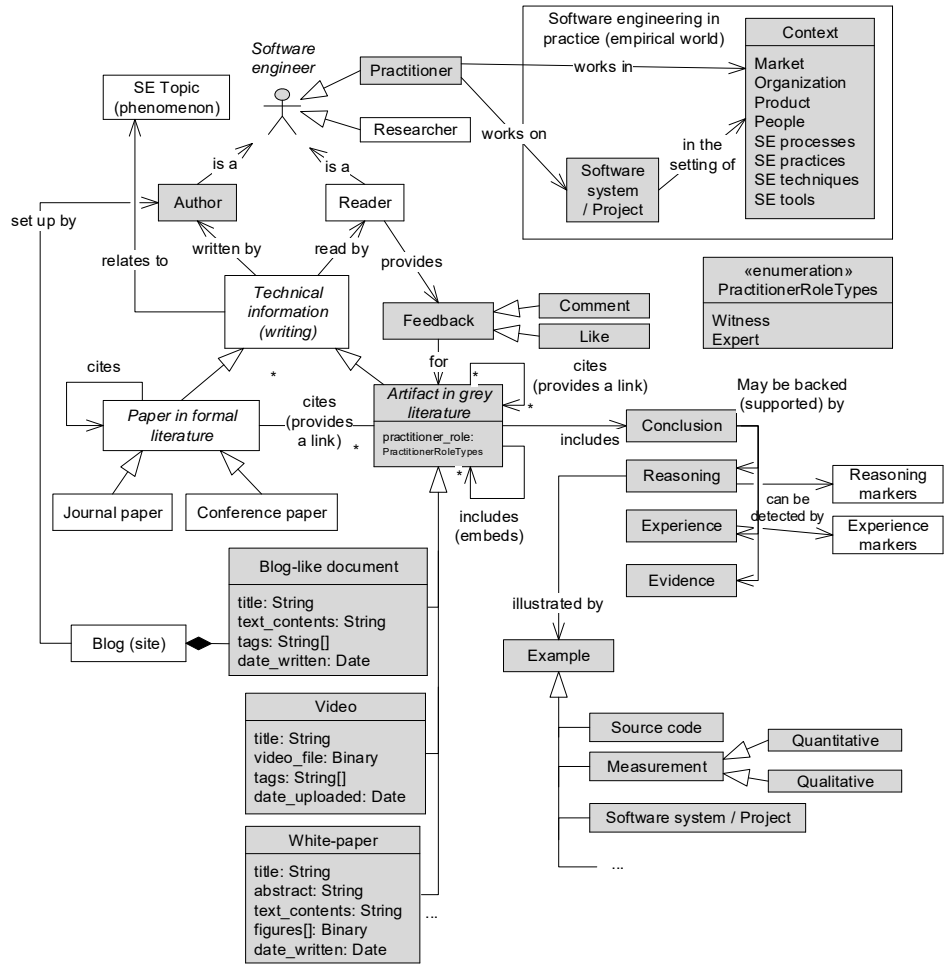

**Figure 2- A context diagram showing the context of GL in SE**

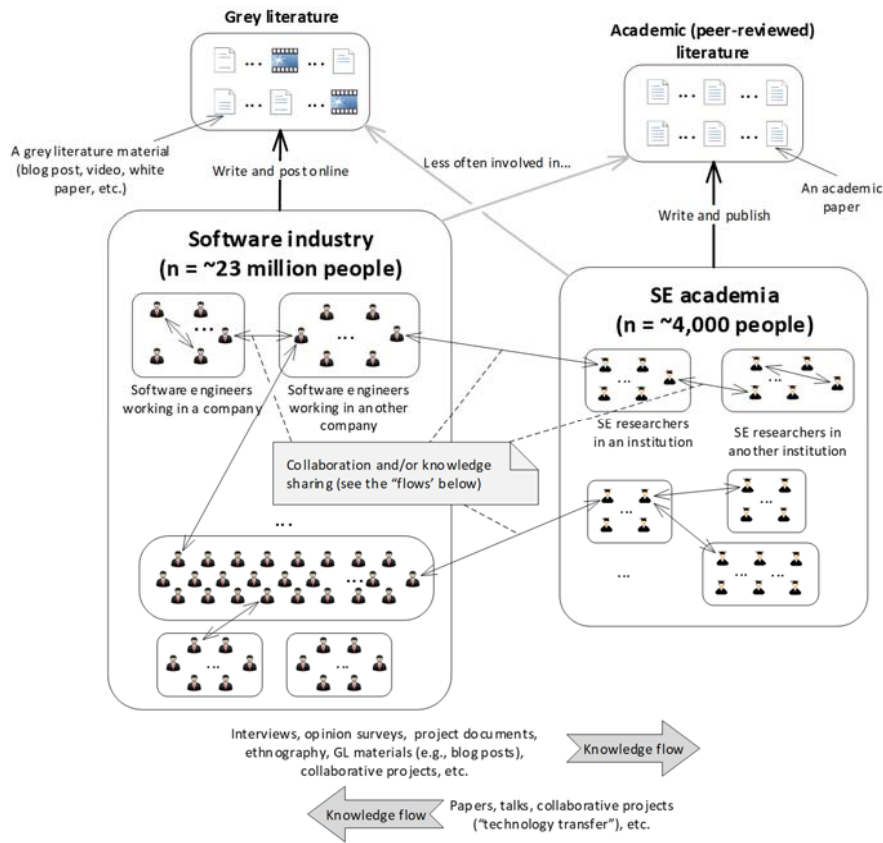

**Figure 3- Visualizing the community of software practitioners and researchers, and the knowledge flow between them (including GL); from Garousi et al. (2018)**

## 3.3. Process of generating the GL content in SE

To better understand the nature of GL in SE, it is also important to characterize the process by which GL content is generated. We depict such a process in Figure 4 which is a simplified version of the ideas presented in Rainer and Williams (2019). The process presented in the figure is intended only as an illustrative example, not as an accurate descriptive account or a prescription of the processes that should occur.

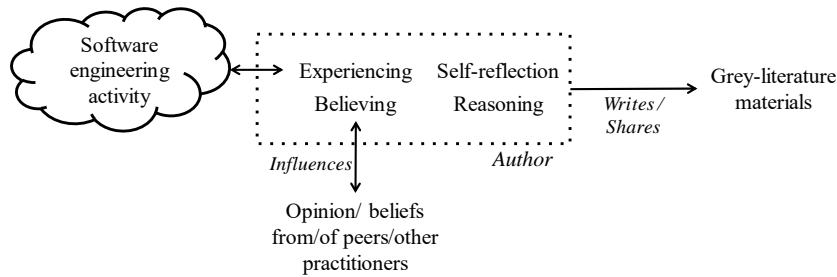

**Figure 4-Process model of the generation of GL contents; simplified from Rainer and Williams (2019)**

In terms of research, we are interested in what the GL author is able to describe of real-world software engineering practice. These descriptions are obviously filtered through the processes that occur between *experiencing* and *reporting* in the model. Many of these processes are internal to the GL author. These internal processes therefore introduce threats to validity relating to subjectivity, and also challenges to research due to the invisibility of these processes. Peer review helps to counteract these threats by independently reviewing the *outputs* from the internal processes, rather than reviewing the processes themselves. Chapter 15 of this book presents important concepts about systematic assessment of threats to validity in software engineering secondary studies, and some of those ideas can be applied when conducing secondary studies involving GL.

In terms of the internal processes represented in Fig. 4:

- Experiencing is an active engagement between the author and the empirical world. Experiencing can take place at different levels of scope and resolution, e.g., directly experiencing programming in contrast to experiencing the 'behavior' of a software organization. The formation of experience is influenced by prior beliefs and in turn influences those beliefs; and is influenced by self-reflection and reasoning.
- Beliefs are defined as conceptions, personal ideologies, worldviews and values that shape practice and orient knowledge. Passos et al. (2011) investigated the role of beliefs in software practice.
- Underpinning the processes that occur within the author, the author has the ability to self-reflect (to some degree) and to reason (to some degree) about her or his experiencing, beliefs and reporting.

Other peoples' beliefs may influence the author. As noted earlier, Devanbu et al. (2016) and Rainer et al. (2003) have investigated the role of others' beliefs. Figure 4 indicates that, finally, the author reports information that may include some description of their experience of software practice, some expression of her or his beliefs, and some degree of reasoning relating these experiences and beliefs. It is very likely that the information reported will be incomplete in some way(s), which could also be the case for papers written by researchers.

### 3.4. GL as a source of knowledge and evidence in SE

An emerging view in SE is that a large amount of SE-related information and experience is becoming available, much of it in the GL, and those data need to be more effectively used to solve practical issues and to push SE research forward, e.g., reported by Garousi et al. (2016a), Rainer (2017), Williams and Rainer (2017), and Rainer and Williams (2019). Lawrence and Giles (1999) observe that this situation occurs in my other disciplines. MacDonald et al. (2007) state that "*the problems of awareness [e.g., for using GL in SE research] persist, even though most of the new information is now digitally produced and arguably easier to access*. It has been recognized in other disciplines that the diffusion, use, and influence of such GL information are complex and variable processes, e.g., by Farace (1997).

GL has already been recognized as a knowledge source in other research areas, e.g., in medicine, with studies by Chavez et al. (2007) and Pappas and Williams (2011); and in earth sciences, with the study by Augusto et al. (2010).

Pappas and Williams (2011) stated that: "*Because of the delay between research and publication, and because of the potential that some important research may never be published, access to innovative information is challenging. Grey literature is a tool to fill that void*".

Garousi et al. (2019) propose one approach to combine knowledge from both GL and published literature: the Multivocal Literature Reviews (MLR). An MLR is a form of a Systematic Literature Review (SLR), which includes the GL in addition to the published literature (e.g., journal and conference papers). MLRs are useful for both researchers and practitioners since they provide summaries of both the state-of-the art and –practice in a given area.

In an MLR on when and what to automate in software testing (abbreviated ManAutoTest MLR in the following) conducted by Garousi and Mäntylä (2016), the researchers reviewed the formal and grey literature. If GL would have been excluded from the pool of papers, a significant body of experience and knowledge from practicing test engineers on the topic would have been missed. To put this in quantitative terms, we partitioned the synthesis of a major output of that MLR (factors to be considered for deciding when and what to automate in testing) by the type of source where they were mentioned in: either formal or GL, as shown in Figure 5. As we can see, out of the total of 15 factor categories, GL sources contributed a total of 219 occurrences (instances) while academic sources discussed only 67 occurrences.

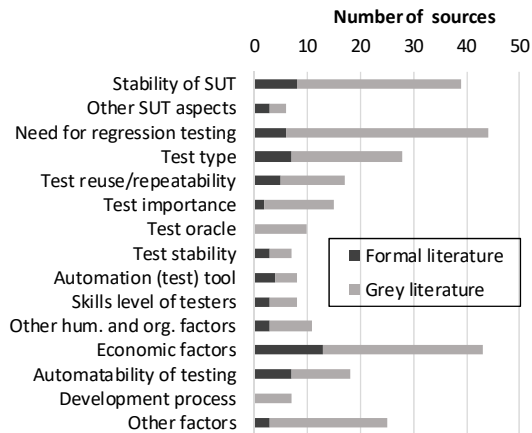

**Figure 5- A main output of the MLR on ManAutoTest; from Garousi and Mäntylä (2016)**

Furthermore, we can see from the figure that, if we were to not include the GL, two categories (*test oracle* and *development process*) would not have been identified in the study. The study demonstrates that GL can be a major source of knowledge and experience. In addition, we extracted in the MLR study a large number of qualitative quotes, related and in support of the factors presented in Figure 5, e.g., a presentation by IBM engineers expressed: "*Main Application has lot of inter-dependency with other Applications which in turn cannot be automated.*", referring to the System Under Test (SUT)-related factors.

Additionally, we found in the ManAutoTest MLR that the type of evidence found in GL were generally either: valid viewpoints, ideas of cause-effect relationships that could be scientifically studied, as well as explanations of why and in what context certain heuristics worked while others did not. We did not however find any sophisticated (hard-core) empirical evidence, such as controlled experiments, in the GL. The stated findings were mostly based on claims and experience. Also, the source of evidence was difficult to identify as the reporting was low quality. Furthermore, we observed in our study that replication of the GL results was not generally possible.

In summary, we can see that the MLR leveraged the readily available GL knowledge on the internet to synthesize the data and answer the important RQs of the MLR study. If no GL data were to be used, the researchers had to conduct interviews, and/or opinion surveys that are often costly and may lead to the same outcomes. Thus, we can see that using GL as a knowledge source can save research costs and also improve research quality.

While GL can be useful as a source of knowledge and evidence in SE research, we raise some caution about how far one can go (scientifically) with GL-based evidence. While such evidence could clearly complement empirical studies in SE, it cannot substitute conventional data gathered in traditional empirical studies. As we will discuss in one of the next sub-sections, there are inherent challenges when using GL in SE research, e.g., the issue of quality-assurance of GL materials. Chapter 16 of this book presents concepts and approaches about evidence aggregation in software engineering, most of which can be applied for aggregation of evidence from GL.

## 3.5. Types of GL in SE

As noted above, there are different types of GL, for example white-papers, blog posts, videos. Within a particular item of GL for SE, there can be considerable variety of content, e.g., a web page can contain text, formatted tables, static images (that may themselves present text or tables), animated images (e.g., GIFs), videos (that again may contain text and tables) and audio. Thus, there is a much greater diversity of types of GL and content within an item of GL compared to academic literature. We list different dimensions of variability in GL materials in Table 2.

**Table 2- Dimensions of variability in GL materials; from Rainer and Williams (2019)**

| Dimension | Explanation and examples |
|---|---|
| Quality of written language | For example, the formality of language. |
| Natural language | Most research appears to focus on English but there are, of course, a very wide range of other languages to consider. |
| Media | Video, Text, Static image, Animated image, Audio, Presentations |
| 'Encoding' of the media | Text with, for example, HTML |
|  | (Proprietary) binary formats e.g., Adobe PDF |
| Structure | Headings, sub-headings |
| Content | Reasoning, e.g., claims, reasons, arguments |
|  | Opinions |
|  | Reporting of actual experience, perhaps as a `war story' |
|  | Code-related information e.g., source code, documentation, API |
|  | Web links e.g., URLs |
|  | (Tables of) data |
|  | Citations |

## 3.6. Scale of GL in SE

It is hard to establish the quantity of general GL available online (without even considering variation in the quality of GL) aside from the challenge of considering the scale of SE-specific GL. Consequently, we briefly report a range of example measures for GL in general and for SE.

*Statista*, an online market research and business intelligence portal, reported in October 2018 that Tumblr, a popular blog platform, alone has 440m blogs[4]. As of January 2019, WordPress self-reported[5] that: "*Users produce about 136.2 million new posts and 77.7 million new comments each month*". None of these statistics relate specifically to GL for SE, and it is unlikely that these statistics would report GL hosted on intranets.

Choi maintains a curated list[6] of blogs focusing on SE, classified by type (i.e., company, individual/group, and product/technology). Choi lists approximately 650 blogs, of which approximately 250 are written by individuals. Panji maintains a curated list of 185 software-related corporate blogs[7], e.g., AutoDesk, BBC, DropBox, Facebook, LinkedIn, Mozilla and NetFlix. Merchant maintains a list of over 50 tech blogs[8]. Abstracta provide a list of 75 blogs and websites on software testing[9]. By contrast, Zalecki maintains a list[10] of software *podcasts*. He states that he is subscribed to over 100 podcasts, although lists eleven at his site. In their systematic GL review of microservices, Soldani *et al*. Soldani et al. (2018) observed a "*... massive proliferation of grey literature [on microservices], with more than 10,000 articles on disparate sub--topics...* ".

### 3.7. Benefits of utilizing the GL in SE research

Rainer and Williams (2019) reviewed research on the benefits, challenges and research directions for the use of blogs in software engineering research. They identified a number of benefits to the use of blogs. Many of these benefits may apply to the use of GL more generally but we focus here on SE. The benefits are summarized in Table 3.

**Table 3- Benefits of utilizing the GL in SE research, based on Rainer and Williams (2019)**

| |
|---|
| In general, GL materials:<br>1. provide information on practitioners' contemporary perspectives on important topics relevant to practice and to research, and<br>2. promote the voice of practitioners |
| In particular, GL materials (such as blog-like documents) provide (access to) information on the practitioner's:<br>1. experience and inexperience of theirs' and others' software practice<br>2. motivations for that practice<br>3. values relating to that practice<br>4. emotions relating to that practice<br>5. beliefs about software practice<br>6. empirical data from their practice, and<br>7. explanations of that practice. |

---

[4] https://www.statista.com/statistics/256235/total-cumulative-number-of-tumblr-blogs/
[5] https://wordpress.com/activity/
[6] https://github.com/kilimchoi/engineering-blogs
[7] https://github.com/sumodirjo
[8] https://github.com/amitmerchant1990/tech-blogs
[9] https://abstracta.us/blog/75-best-software-testing-blogs/
[10] https://michalzalecki.com/curated-list-of-podcasts-for-software-developers/

| |
|---|
| In providing such information, GL materials: <br> 1. help bridge the divide between research and practice <br> 2. complement the research literature by 'filling in gaps' in research, and <br> 3. help to counteract bias findings, as a result of publication bias in the research literature. |
| GL materials should be considered when Williams and Rainer (2017): <br> 1. the topic of the research is complex <br> 2. the topic is not 'solvable' by using only the peer--reviewed research literature <br> 3. there is a lack of quantity and/or quality of best evidence from research, or a lack of consensus in the research <br> 4. context is important to the study of the topic <br> 5. the researcher intends to challenge existing assumptions and findings, either in research or practice, or both <br> 6. a synthesis of practice and research would be valuable to either or both communities; <br> 7. the researcher intends to consider trends over time, and <br> 8. the researcher seeks to better understand, assess or demonstrate the impact of research in relation to a particular topic. |
| Methodologically, the use of GL materials in research helps researchers to: <br> 1. assess and address publication bias <br> 2. compensate for the (un)availability of other sources of evidence <br> 3. increase research visibility into actual software practice <br> 4. access harder-to-access practitioners e.g. due to logistics, or demographics <br> 5. gather information for the research in a non-invasive way <br> 6. scale-up their research to, or with, larger samples <br> 7. complement and triangulate with, other sources of data <br> 8. provide an audit trail of their research, and <br> 9. replicate each other's study through public access to original data. |

## 3.8. Challenges of using GL in SE research

As well as the benefits of using blog posts (identified earlier), Rainer and Williams (2019) also identified a number of challenges to the use of blogs as a type of GL. These challenges were organized into several themes and are summarized here in Table 4.

**Table 4- Challenges of working with and using GL in SE research, based on Rainer and Williams (2019)**

| Challenges themes | Concrete challenges |
|---|---|
| Foundations e.g. there are a lack of… | • Formal definitions of GL and GL materials <br> • Formal models of GL materials and content, in particular; <br>   o a data model of GL materials and content; an <br>   o a process model of the creation, review and publication of GL materials and content; <br> • Frameworks for evaluating the quality of GL materials and content, and classifying those materials and content; |
| Inherent nature of GL materials | There are challenges managing… <br> • The very large quantity of GL materials <br> • The variability of GL materials <br> • The uncertain process for generating, publishing and revising the content of GL materials |
| Resources | There are a lack of… <br> • Central repositories of GL materials; <br> • Tools to work with GL materials and content, for example: <br>   o to select the higher-quality documents when performing a search; and |
| | ○ to select particular types of GL materials e.g. those reporting experience, values, explanations etc.<br>• Datasets and corpora of GL materials |
| Quality-assurance | While some efforts have started, e.g. Garousi et al. (2019), there is a shortage of:<br>• Well-developed and accepted checklists for the quality assurance of various aspects of GL materials including:<br>○ the author;<br>○ the document;<br>○ the content of the document e.g. claims;<br>○ the readers' assessment of the credibility of the document;<br>○ the readers;<br>○ the readers' feedback on the document e.g. comments, shares, up-votes; |
| Methodology | • The evidential value of blog-like content;<br>• The appropriate research methods to use with GL materials and content. |

## 3.9. Diversity in quality and degree of evidence in GL materials

Since processes for GL are more diverse and less controlled, compared to academic literature, the quality of GL is more diverse and often more difficult to assess. The quality of GL determines whether data from GL or conclusions raised in GL can be used and analyzed (see Sect. 4). Garousi et al. (2019) compiled a quality assessment checklist for GL shown in Table 5. It contains the criteria of, authority of the producer, methodology, objectivity, date, position with respect to related sources, novelty, impact, and outlet type as well as assessment questions for each criterion.

**Table 5- Quality assessment checklist for GL in SE**

| Criteria | Questions |
|---|---|
| Authority of the producer | • Is the publishing organization reputable? E.g., the Software Engineering Institute (SEI)<br>• Is an individual author associated with a reputable organization?<br>• Has the author published other work in the field?<br>• Does the author have expertise in the area? (e.g. job title principal software engineer) |
| Methodology | • Does the source have a clearly stated aim?<br>• Does the source have a stated methodology?<br>• Is the source supported by authoritative, contemporary references?<br>• Are any limits clearly stated?<br>• Does the work cover a specific question?<br>• Does the work refer to a particular population or case? |
| Objectivity | • Does the work seem to be balanced in presentation?<br>• Is the statement in the sources as objective as possible? Or, is the statement a subjective opinion?<br>• Is there a vested interest? E.g., a tool comparison by authors that are working for a particular tool vendor<br>• Are the conclusions supported by the data? |

18 Benefitting from the grey literature in software engineering research| Date | • Does the item have a clearly stated date? |
|---|---|
| Position w.r.t. related sources | • Have key related GL or formal sources been linked to / discussed? |
| Novelty | • Does it enrich or add something unique to the research?<br>• Does it strengthen or refute a current position? |
| Impact | • Normalize all the following impact metrics into a single aggregated impact metric (when data are available): Number of citations; Number of backlinks; Number of social media shares (the so-called "alt-metrics"); Number of comments posted for a specific online entry, like a blog post or a video; Number of page or paper views. |
| Outlet type | • 1st tier GL (measure=1): High outlet control/ High credibility: Books, magazines, theses, government reports, white papers<br>• 2nd tier GL (measure=0.5): Moderate outlet control/ Moderate credibility: Annual reports, news articles, presentations, videos, Q/A sites (such as StackOverflow), Wiki articles<br>• 3rd tier GL (measure=0): Low outlet control/ Low credibility: Blogs, emails, tweets |

For each type of GL, the relevant quality criteria have to be selected, adapted and finally assessed, which can for instance be done on a two-point Likert scale with values "yes" or "no"; see for example, Garousi et al. (2019). For instance, the number of online comments to measure the impact only exist for source types open for comments like blog posts, news articles or videos. A highly commented blog post may indicate popularity, but on the other hand, spam comments may bias the number of comments, thus invalidating the high popularity.

## 4. How GL can be used / analyzed in SE

The SE research community has started to use the information and evidence from the GL in different ways.

### 4.1. Review of how GL has been used / analyzed in SE research

Table 6 classifies the SE research community's use of GL. We distinguish in Table 6 different ways of utilizing / analyzing the GL in SE research. The first three types of study concern the use of GL in a primary study, ranging from studies with a specific focus on GL to those studies that only cite GL. The fourth type of study concerns secondary studies i.e., the systematic review of GL. We discuss these types in more detail in the following subsections.

**Table 6- Different ways of utilizing / analyzing the GL in SE research community so far**

| Study types | Type of usage / analysis | Example papers |
|---|---|---|
| Primary studies | Analyzing GL materials with qualitative approach | • Using argumentation theory to analyze software practitioners' defeasible evidence, inference and belief Rainer (2017) |
| (from specific-focus on GL to only citing GL) | | • An analysis of major pivots of software start-ups Bajwa et al. (2017)<br>• Analyzing the motivations and challenges of developers for blogging Parnin et al. (2013) |
| | Analyzing GL with quantitative approach | • Measuring API documentation: 1,730 websites and 376 blog posts Parnin and Treude (2011)<br>• What are mobile developers asking about? a large scale study using stack overflow Rosen and Shihab (2016) |
| | Citation to GL: GL materials are cited in research papers as related works / examples. | Many papers in SE cite GL materials for different reasons, e.g., to motivate the papers. Two examples of widely-cited GL materials in SE are:<br><br>• The economic impacts of inadequate infrastructure for software testing Planning (2002)<br>• Various editions of the Standish Group's "Chaos" report The Standish Group (2019) |
| Secondary studies | Systematic reviews involving GL | • An MLR on iOS applications testing Kulesovs (2015)<br>• A GLR on choosing the right test automation tool Raulamo et al. (2017)<br>• An MLR on when to automate in testing Garousi and Mäntylä (2016) |

### 4.1.1. Qualitative analysis of GL materials

Many SE researchers are analyzing GL material and answering GL specific research questions even when they do not explicitly acknowledge it. The grey literature can be analyzed both with qualitative and quantitative methods (see next section). With qualitative methods we mean analysis methods where humans read, analyze, and classify GL text in order to produce knowledge. When using a qualitative approach, one can use approaches presented in qualitative research guideline books and articles, e.g., those by Patton (2002) and Cruzes and Dyba (2011).

We can find qualitative works in this area. In some papers, humans analyze and classify the particular GL contents and explore motivations for GL production in software development. Parnin et al. (2013) analyzed why and about what the software developers write blogs. They found that the blogs covered multiple topics such as code and tool tutorials, new releases and enhancement to the products the developers were working with, and general technology discussions for example. Blogging was motivated by personal branding, evangelism and getting feedback, and finally for personal knowledge repository. A study of similar nature was later conducted by MacLeod et al. (2015) on software developers' YouTube videos that found video content was more about technical topics such as development experience, implementation choices and data structures. Videos were also seen as an alternative to blogging and many similar motivations for video creations existed as for blogging.

Bajwa et al. (2017) use GL of software start-ups and analyze their business pivots. The authors frame their study as a case study on secondary data that the authors collected from various websites. They find that software start-ups pivot for 14 reasons (triggers) such as negative customer reaction, unable to beat a competitor, and technological challenge. They also find evidence of ten different pivot types such as switching to a different problem and zoom-in where a particular feature becomes the whole product.

There are other studies that use a set of GL materials, but do not survey a large/r set of literature, instead adopting a kind of case-study approach. As one example, Rainer (2017) used argumentation schemes to qualitatively analyze information reported by Joel Spolsky in one of his blog posts, entitled *The Language Wars*. Rainer formally modelled the integration of argumentation structures and professional experience, and then showed how the arguments and experience can be related to previous research. Rainer's in-depth analyses of one blog post may be understood as a case study to complement the survey-like studies of MLRs and Grey Literature Reviews (GLRs).

### 4.1.2. Quantitative analysis of GL materials

In the quantitative analysis of GL, research methods range from simple frequency counting to advanced machine learning used for natural language processing such as Rosen and Shihab (2016) topic modeling LDA (Latent Dirichlet Allocation) and Efstathiou et al. (2018) use of word embeddings. Much of the quantitative analyses of GL appears to concentrate around a small number of sources, principally Stack Overflow. This may be because the data is easy to access and relatively well structured. But in addition to StackOverflow, one can find more analyses of blogs, e.g. by Parnin and Treude (2011), and emails, e.g., by Sharma et al. (2017), that should offer a more multivocal view. Next, we present a few examples of studies using quantitative analysis of GL.

Gruetze et al. (2016) examined topic shifts, by analyzing tags in software development QA-site StackOverflow. The authors showed declining trends in tags like Delphi and Database-Design while increasing trends were found for the programming language "R" and Node.js for example. They also show that automated tagging of posts is improved when the time of the post is considered.

As noted already, Rosen and Shihab (2016) also analyzed StackOverflow but with respect to questions that mobile developers are asking about. They analyse 13 million posts and use LDA to cluster the data. They find that mobile developers ask about "app distribution, mobile APIs, data management, sensors and context, mobile tools, and user interface development".

Quantitative analyses of GL in StackOverflow have also been used to create guidelines of how to create good StackOverflow posts, e.g., by Calefato et al. (2018). The authors suggest using quantitative analysis that successful StackOverflow questions are: short, have code snippets, do not abuse uppercase letter and have neutral emotional expressions. So the analysis of GL can be used to provide advice on how to write better GL.

Sharma et al. (2017) performed quantitative analysis on email discussion in Python language evolution. They collect a data set of over 40,000 emails. They found that technical discussion receives clearly the highest volume of emails over social and process issues. The authors conclude that this shows that Python developers mostly care and are passionate about technical features of the language.

### 4.2. Citations to GL in SE papers

Many papers in SE cite GL materials for different reasons, e.g., to motivate the papers, to use their insights/data, etc. Two example widely-cited GL materials in SE are: (1) A technical report entitled "*The economic impacts of inadequate infrastructure for software testing*", conducted by Research Triangle Institute (2002), for the American National Institute of Standards and Technology (NIST), which was cited about 700 times accordingly to Google Scholar (October 2019); and (2) Various editions of the Standish Group's "Chaos" report, e.g., The Standish Group (2019).

We believe that by providing more citations and getting insights form GL in research papers, researchers will contribute to a stronger linkage between industry and academia, as mentioned in Garousi et al. (2016b) and Garousi et al. (2017a), since readers and follow-up research studies will be encouraged to use more real-world industrial approaches and data.

### 4.3. Systematic reviews using GL sources

Systematic reviews systematically select, review, and synthesize knowledge in a given topic of SE. Traditionally, since the inception of Evidence-Based Software Engineering (EBSE) by Kitchenham et al. (2004), two types of review studies have been published in the SE community: Systematic Literature Reviews (SLR), and Systematic Literature Mappings (SLM or SM).

With more awareness for GL in SE, recent review studies in SE have started to include GL, e.g., Garousi et al. (2019). We first discuss the different types of systematic reviews which include GL and then discuss the guidelines to conduct such studies.

To include GL, four new types of review studies have emerged, as discussed by Garousi et al. (2019): (1) Multivocal Literature Review (MLR), (2) Multivocal Literature Mapping (MLM), (3) Grey Literature Mapping (GLM), and (4) Grey Literature Review (GLR). An MLR is a form of an SLR which includes the GL in addition to the published literature. To clearly distinguish all different types of review studies in SE, we depict the relationship among them in Figure 6.

As we see in Figure 6, the differentiation factors of six types of systematic secondary studies are: types of analysis, and types of sources under study. For example, the difference between an MLR and a GLR is that, while the former reviews both GL and published literature, the latter reviews only the GL. The difference between an MLM and an MLR is that, while both analyze GL, the former reviews only classified the pool of sources, the latter synthesizes the evidence from those sources in addition.

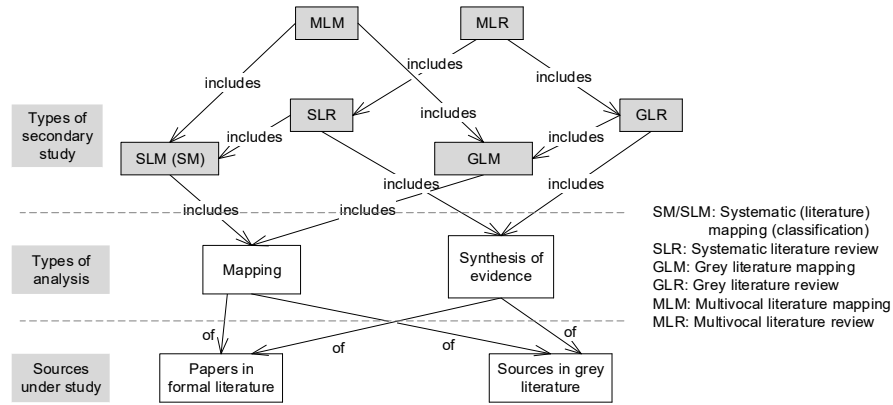

**Figure 6-Relationship among different types of systematic secondary studies (from Garousi et al. (2019))**

We looked at recent review studies in SE involving GL. We were able to find 18 such studies as shown in Table 7. Note that this list only contains the review studies involving GL focusing on SE. There have also been a recent trend on review studies involving GL in other areas of CS, e.g., an MLR on server-less computing by Sadaqat et al. (2018).

As highlighted in the table, the authors of this chapter have been involved in six (6) of these studies. As it can be seen in the table, there has been a *sharp* increase in such studies in recent years, as 9 out of 18 papers were published in 2018.

For each of the studies in Table 7, we also show the number of academic literature (AL) sources, number of GL sources, and percentage of GL sources reviewed in that study. Needless to say, the ratio would be %100 for GLR studies. The ratios, in a sense, denote the scale of AL versus GL knowledge in a given topic. For example, for the topic of involving security in DevOps (DevSecOps), the numbers of AL/ GL sources are 2/50 (a GL ratio of 96%), while in the topic of ethics in requirements engineering, the numbers of AL/ GL sources are 98/34 (a GL ratio of 26%).

**Table 7-A summary of the recent review studies involving GL**

| Review topic | Type | | Year | Ref. | # of AL sources | # of GL sources | % of GL sources |
| --- | --- | --- | --- | --- | --- | --- | --- |
| | MLR | GLR | | | | | |
| Technical debt | x | | 2013 | Tom et al. (2013) | 0 | 35 | 100% |
| iOS applications testing | x | | 2015 | Kulesovs (2015) | 12 | 9 | 42% |
| When to automate in testing | x | | 2016 | Garousi and Mäntylä (2016) | 26 | 52 | 66% |
| Gamification of SW testing | x | | 2016 | Mäntylä and Smolander (2016) | 6 | 14 | 70% |
| Relationship of DevOps to agile | x | | 2016 | Lwakatare et al. (2016) | 33 | 201 | 86% |

| | | | | | | |
|---|---|---|---|---|---|---|
| Characterizing DevOps | x | | 2016 | Franca et al. (2016) | 24 | 19 | 44% |
| Test maturity and test process improvement | x | | 2017 | Garousi et al. (2017b) | 130 | 51 | 28% |
| Involving security in DevOps (DevSecOps) | x | | 2017 | Myrbakken and Colomo-Palacios (2017) | 2 | 50 | 96% |
| Choosing the right test automation tool: a GLR | | x | 2017 | Raulamo et al. (2017) | 0 | 53 | 100% |
| Smells in SW test code | x | | 2018 | Garousi and Küçük (2018) | 46 | 120 | 28% |
| Serious games for SW process | x | | 2018 | Calderón et al. (2018) | 6 | 1 | 14% |
| Pains and gains of micro-services | | x | 2018 | Soldani et al. (2018) | 0 | 51 | 100% |
| Relevance of software engineering research | x | | 2018 | Garousi et al. (2018) | 33 | 13 | 28% |
| Ethics in requirements engineering | x | | 2018 | Aberkane (2018) | 98 | 34 | 26% |
| Function-as-a-Service software development | | x | 2018 | Leitner et al. (2018) | 0 | 50 | 100% |
| Adopting the Scaled Agile Framework (SAFe) | x | | 2018 | Putta et al. (2018) | 52 | 47 | 47% |
| Monolithic repositories (Monorepos) | x | | 2018 | Brito et al. (2018) | 2 | 21 | 91% |
| Use of DevOps for e-Learning systems | x | | 2018 | Sánchez-Gordón and Colomo-Palacios (2018) | 3 | 22 | 88% |

Grey literature, and grey literature reviews, inevitably have their limitations. Rainer and Williams (2019) identified several challenges with using blog posts in software engineering research. Many of these challenges apply to GL e.g., the vast quantity of GL available, and the variability in the quality of GL. MLRs are one approach to addressing the limitations of GL i.e., by combining GL with AL. As researchers conduct more reviews using GL so the community can develop better guidelines, checklists and methodology for using GL in research.

## 5. Recommended Further Reading

The usage of grey literature (GL) in software engineering is strongly related to evidence-based methods and literature reviews in software engineering.

Kitchenham et al. (2015) provided a comprehensive book on evidence-based software engineering and systematic reviews.

For the main types of systematic literature studies in software engineering, i.e., systematic literature reviews and mapping studies, there are highly referenced guideline papers, such as: guidelines by Kitchenham and Charters (2007) for Systematic Literature Reviews (SLRs), which are also extensively discussed together with background information in the aforementioned book Kitchenham et al. (2015); and Petersen et al. (2015)'s guidelines for Systematic Mapping Studies (SMSs). You could perhaps include a citation to Rapid Reviews?

However, none of the above guidelines explicitly discuss GL. Garousi et al. (2019) filled this gap and provided guidelines for including GL and conducting Multivocal Literature Reviews (MLRs) in software engineering. Researchers are encouraged to consult those guidelines when planning MLR or other types of studies involving GL. The guidelines for MLRs in SE cover planning, conducting and reporting the review. The step on conducting an MLR comprises guidelines for the search process, source selection, study quality assessment, data extraction and data synthesis.

Marsolek et al. (2018) provided an overview of the usage of GL in other fields arts, business, education, health sciences, humanities, multidisciplinary research, natural sciences, physical sciences and engineering, and social sciences. Especially, in health sciences, GL and its analysis is well established and there is even a book by Bonato (2018) on searching the GL.

## 6. Conclusion

The goal of this chapter has been to provide an overview to GL in SE, together with insights on how SE researchers can effectively use and benefit from the knowledge and evidence available in the vast amount of GL. We first reviewed the general concept of GL and provide background information. We then discussed the state of GL in SE research, including context, types, diversity and scale of GL in SE research and practice. We then proposed and discussed five approaches for using and analyzing GL in SE research: (1) Analyzing GL materials to answer GL-specific RQs; (2) Using certain GL materials for qualitative studies; (3) Using certain GL materials quantitative studies; (4) Citing GL materials; and (5) Systematic reviews involving GL.

As discussed above and also as indicated in other studies, e.g., by University of New England (2019), the reality is that researchers mostly write for, and read from, scientific papers published in the academic, peer-reviewed literature; and by contrast, practitioners mostly write for, and read from, materials published in the grey literature. By reviewing how GL has been used in SE research, this chapter aims to encourage further use of GL in SE research. We recommend all SE researchers to reduce the gap between academia and industry via using GL materials in the five forms as discussed in this chapter.

# References

Aberkane, A. (2018). "Exploring Ethics in Requirements Engineering." Master thesis, Utrecht University.

Adams, R. J., P. Smart and A. S. Huff (2016). "Shades of Grey: Guidelines for Working with the Grey Literature in Systematic Reviews for Management and Organizational Studies." International Journal of Management Reviews: n/a-n/a.

Augusto, L., M. R. Bakker, C. Morel, et al. (2010). "Is 'grey literature' a reliable source of data to characterize soils at the scale of a region? A case study in a maritime pine forest in southwestern France." European Journal of Soil Science 61(6): 807-822.

Bajwa, S. S., X. Wang, A. Nguyen Duc, et al. (2017). ""Failures" to be celebrated: an analysis of major pivots of software startups." Empirical Software Engineering 22(5): 2373-2408.

Banks, M. A. (2006). "Towards a continuum of scholarship: The eventual collapse of the distinction between grey and non-grey literature." Publishing Research Quarterly 22(1): 4-11.

Bansal, N., F. Chiang, N. Koudas, et al. (2007). Seeking Stable Clusters in the Blogosphere, VLDB Endowment, 806–817.

Bonato, S. (2018). Searching the Grey Literature: A Handbook for Searching Reports, Working Papers, and Other Unpublished Research, RL & Associates LLC.

Briand, L. (2012). "Embracing the Engineering Side of Software Engineering." IEEE Software 29(4): 96-96.

Brito, G., R. Terra and M. T. Valente (2018). "Monorepos: A Multivocal Literature Review." arXiv preprint arXiv:1810.09477.

Brooks, F. (1995). The Mythical Man-Month: Essays on Software Engineering, Pearson Education.

Calderón, A., M. Ruiz and R. V. O'Connor (2018). "A multivocal literature review on serious games for software process standards education." Computer Standards & Interfaces 57: 36-48.

Calefato, F., F. Lanubile and N. Novielli (2018). "How to ask for technical help? Evidence-based guidelines for writing questions on Stack Overflow." Information and Software Technology 94: 186-207.

Chavez, T. A., A. H. Perrault, P. Reehling, et al. (2007). "The impact of grey literature in advancing global karst research: an information needs assessment for a globally distributed interdisciplinary community." Publishing Research Quarterly 23.

Corporation, E. D. (2018). "Global Developer Population and Demographic Study 2018." https://evansdata.com/reports/viewRelease.php?reportID=9.

Cruzes, D. S. and T. Dyba (2011). Recommended steps for thematic synthesis in software engineering. International Symposium on Empirical Software Engineering and Measurement, IEEE, 275-284.

DeMarco, T. and T. Lister (2013). Peopleware: productive projects and teams, Addison-Wesley.

Devanbu, P., T. Zimmermann and C. Bird (2016). Belief & evidence in empirical software engineering. IEEE/ACM International Conference on Software Engineering, IEEE, 108-119.

Efstathiou, V., C. Chatzilenas and D. Spinellis (2018). Word embeddings for the software engineering domain. Proceedings of the International Conference on Mining Software Repositories, ACM, 38-41.

Elliott, C. (2019). "Jinfo Blog: Garner Additional Research Sources with Grey Literature." https://web.jinfo.com/go/blog/70203.

Facca, F. M. and P. L. Lanzi (2005). "Mining interesting knowledge from weblogs: a survey." Data & Knowledge Engineering 53(3): 225-241.

26      Benefitting from the grey literature in software engineering researchFarace, D. J. (1997). "Rise of the phoenix: A review of new forms and exploitations of grey literature." Publishing Research Quarterly **13**(2): 69-76.

Franca, B. B. N. d., J. Helvio Jeronimo and G. H. Travassos (2016). Characterizing DevOps by Hearing Multiple Voices. Proceedings of the Brazilian Symposium on Software Engineering, 53-62.

Garousi, V., M. Borg and M. Oivo (2018). "Cut to the chase: Revisiting the relevance of software engineering research." arXiv preprint arXiv:1812.01395.

Garousi, V., M. Felderer, J. M. Fernandes, et al. (2017a). Industry-academia collaborations in software engineering: An empirical analysis of challenges, patterns and anti-patterns in research projects. Proceedings of International Conference on Evaluation and Assessment in Software Engineering, Karlskrona, Sweden, 224-229.

Garousi, V., M. Felderer and T. Hacaloğlu (2017b). "Software test maturity assessment and test process improvement: A multivocal literature review." Information and Software Technology **85**: 16–42.

Garousi, V., M. Felderer and M. V. Mäntylä (2016a). The need for multivocal literature reviews in software engineering: complementing systematic literature reviews with grey literature. International Conference on Evaluation and Assessment in Software Engineering, Limmerick, Ireland, 171-176.

Garousi, V., M. Felderer and M. V. Mäntylä (2019). "Guidelines for including grey literature and conducting multivocal literature reviews in software engineering." Information and Software Technology **106**: 101-121.

Garousi, V. and B. Küçük (2018). "Smells in software test code: A survey of knowledge in industry and academia." Journal of Systems and Software **138**: 52-81.

Garousi, V. and M. V. Mäntylä (2016). "When and what to automate in software testing? A multivocal literature review." Information and Software Technology **76**: 92-117.

Garousi, V., K. Petersen and B. Özkan (2016b). "Challenges and best practices in industry-academia collaborations in software engineering: a systematic literature review." Information and Software Technology **79**: 106–127.

Giustini, D. and D. Thompson (2010). "Finding the Hard to Finds: Searching for Grey (Gray) Literature." https://blogs.ubc.ca/dean/2010/02/finding-the-hard-to-finds-searching-for-grey-gray-literature-2010/.

Gruetze, T., R. Krestel and F. Naumann (2016). Topic shifts in stackoverflow: Ask it like socrates. International Conference on Applications of Natural Language to Information Systems, Springer, 213-221.

Institute for Work & Health (2019). "What researchers mean by... Grey literature." https://www.iwh.on.ca/what-researchers-mean-by/grey-literature.

Inui, K., S. Abe, K. Hara, et al. (2008). Experience mining: Building a large-scale database of personal experiences and opinions from web documents, IEEE Computer Society, 314–321.

Kitchenham, B., D. Budgen and P. Brereton (2015). Evidence-Based Software engineering and systematic reviews, CRC Press.

Kitchenham, B. and S. Charters (2007). Guidelines for Performing Systematic Literature Reviews in Software engineering. EBSE Technical Report. **EBSE-2007-01**.

Kitchenham, B. A., T. Dyba and M. Jorgensen (2004). Evidence-Based Software Engineering. Proceedings of International Conference on Software Engineering, 273-281.

Kulesovs, I. (2015). iOS Applications Testing. Proceedings of the International Scientific and Practical Conference, 138-150.

Kurashima, T., K. Fujimura and H. Okuda (2009). Discovering association rules on experiences from large-scale blog entries. European Conference on Information Retrieval, 546–553.

Kurashima, T., T. Tezuka and K. Tanaka (2006). Mining and visualizing local experiences from blog entries. International Conference on Database and Expert Systems Applications, Springer, 213–222.

Lakshmanan, G. and M. Oberhofer (2010). "Knowledge Discovery in the Blogosphere: Approaches and Challenges." IEEE Internet Computing **14**(2): 24-32.

Lawrence, S. and C. L. Giles (1999). "Accessibility of information on the web." Nature **400**: 107.

Lefebvre, C., E. Manheimer and J. Glanville (2008). Searching for studies. Cochrane handbook for systematic reviews of interventions. J. P. T. Higgins and S. Green, Chichester: Wiley-Blackwell.

Leitner, P., E. Wittern, J. Spillner, et al. (2018). "A mixed-method empirical study of Function-as-a-Service software development in industrial practice." Journal of Systems and Software.

Luzi, D. (2000). "Trends and evolution in the development of grey literature: a review." International Journal on Grey Literature **1**(3): 106-117.

Lwakatare, L. E., P. Kuvaja and M. Oivo (2016). Relationship of DevOps to Agile, Lean and Continuous Deployment: A Multivocal Literature Review Study. Proceedings of International Conference on Product-Focused Software Process Improvement, 399-415.

MacDonald, B. H., R. E. Cordes and P. G. Wells (2007). Assessing the Diffusion and Impact of Grey Literature Published by International Intergovernmental Scientific Groups: The Case of the Gulf of Maine Council on the Marine Environment. Proceedings of the International Conference on Grey Literature, 84-94.

MacLeod, L., M.-A. Storey and A. Bergen (2015). Code, camera, action: How software developers document and share program knowledge using YouTube. Proceedings of the IEEE International Conference on Program Comprehension, 104-114.

Mäntylä, M. V. and K. Smolander (2016). Gamification of Software Testing - An MLR. International Conference on Product-Focused Software Process Improvement, 611-614.

Marsolek, W., K. Cooper, S. Farrell, et al. (2018). "The Types, Frequencies, and Findability of Disciplinary Grey Literature within Prominent Subject Databases and Academic Institutional Repositories." Journal of Librarianship and Scholarly Communication **6**(1).

McKimmie, T. and J. Szurmak (2002). "Beyond Grey Literature: How Grey Questions Can Drive Research." Journal of Agricultural & Food Information **4**(2): 71-79.

Molléri, J. S., K. Petersen and E. Mendes (2016). Survey Guidelines in Software Engineering: An Annotated Review. Proceedings of the ACM/IEEE International Symposium on Empir-ical Software Engineering and Measurement, 58:51–58:56.

Myrbakken, H. and R. Colomo-Palacios (2017). DevSecOps: A Multivocal Literature Review. Conf. on Software Process Improvement and Capability Determination, 17-29.

Pappas, C. and I. Williams (2011). "Grey literature: its emerging importance." Journal of Hospital Librarianship **11**.

Park, K. C., Y. Jeong and S. H. Myaeng (2010). Detecting experiences from weblogs. Proceedings of the Annual Meeting of the Association for Computational Linguistics, 1464–1472.

Parnin, C. and C. Treude (2011). Measuring API documentation on the web, ACM, 25–30.

Parnin, C., C. Treude and M.-A. Storey (2013). Blogging developer knowledge: Motivations, challenges, and future directions. IEEE International Conference on Program Comprehension, 211-214.

Passos, C., A. P. Braun, D. S. Cruzes, et al. (2011). Analyzing the impact of beliefs in software project practices. International Symposium on Empirical Software Engineering and Measurement, IEEE, 444-452.

Patton, M. Q. (2002). "Qualitative research and evaluation methods. Thousand Oaks." Cal.: Sage Publications.

Petersen, K., S. Vakkalanka and L. Kuzniarz (2015). "Guidelines for conducting systematic mapping studies in software engineering: An update." Information and software technology **64**: 1-18.

Planning, S. (2002). "The economic impacts of inadequate infrastructure for software testing." National Institute of Standards and Technology.

Putta, A., M. Paasivaara and C. Lassenius (2018). Benefits and Challenges of Adopting the Scaled Agile Framework (SAFe): Preliminary Results from a Multivocal Literature Review. International Conference on Product-Focused Software Process Improvement, Springer, 334-351.

Rainer, A. (2017). "Using argumentation theory to analyse software practitioners' defeasible evidence, inference and belief." Information and Software Technology **87**: 62-80.

Rainer, A., T. Hall and N. Baddoo (2003). Persuading developers to" buy into" software process improvement: a local opinion and empirical evidence. International Symposium on Empirical Software Engineering, IEEE, 326-335.

Rainer, A. and A. Williams (2019). "Using blog-like documents to investigate software practice: Benefits, challenges, and research directions." Journal of Software: Evolution and Process, In Press.

Raulamo, P., M. V. Mäntylä and V. Garousi (2017). Choosing the right test automation tool: a grey literature review. International Conference on Evaluation and Assessment in Software Engineering, Karlskrona, Sweden, 21-30.

Research Triangle Institute (2002). "The economic impacts of inadequate infrastructure for software testing." American National Institute of Standards and Technology (NIST), Technical report 7007.011.

Rincon, P. (2010). "Stricter checks for climate body." https://www.bbc.com/news/science-environment-11131897.

Rosen, C. and E. Shihab (2016). "What are mobile developers asking about? a large scale study using stack overflow." Empirical Software Engineering **21**(3): 1192-1223.

Sadaqat, M., R. Colomo-Palacios and L. E. S. Knudsen (2018). Serverless computing: a multivocal literature review. Norwegian conference for organizations' use of information technology (NOKOBIT),

Sánchez-Gordón, M. and R. Colomo-Palacios (2018). A Multivocal Literature Review on the use of DevOps for e-Learning systems. Proceedings of International Conference on Technological Ecosystems for Enhancing Multiculturality, 883-888.

Sharma, P., B. T. R. Savarimuthu, N. Stanger, et al. (2017). Investigating developers' email discussions during decision-making in Python language evolution. Proceedings of International Conference on Evaluation and Assessment in Software Engineering. Karlskrona, Sweden, ACM**:** 286-291.

Sharp, H., C. deSouza and Y. Dittrich (2010). Using ethnographic methods in software engineering research. ACM/IEEE International Conference on Software Engineering, 491-492.

Soldani, J., D. A. Tamburri and W.-J. Van Den Heuvel (2018). "The pains and gains of microservices: A Systematic grey literature review." Journal of Systems and Software **146**: 215-232.

Storey, M.-A., L. Singer, B. Cleary, et al. (2014). The (r) evolution of social media in software engineering. Proceedings of the on Future of Software Engineering, ACM, 100-116.

Swanson, R., E. Rahimtoroghi, T. Corcoran, et al. (2014). Identifying Narrative Clause Types in Personal Stories. Proceedings of the Annual Meeting of the Special Interest Group on Discourse and Dialogue, 171–180.

The Standish Group (2019). "CHAOS Report." https://www.standishgroup.com/outline.

Thompson, L. (2001). "Grey Literature in Engineering." Science & Technology Libraries **19**(3-4): 57-73.

Tom, E., A. Aurum and R. Vidgen (2013). "An exploration of technical debt." Journal of Systems and Software **86**(6): 1498-1516.

University of New England (2019). "Grey literature." https://www.une.edu.au/library/support/eskills-plus/research-skills/grey-literature.

Williams, A. and A. Rainer (2017). Toward the use of blog articles as a source of evidence for software engineering research. Proceedings of the International Conference on Evaluation and Assessment in Software Engineering, 280-285.